\documentstyle{article}

\newcommand{\sss}{{\rm s}^2}
\newcommand{\ccc}{{\rm c}^2}
\newcommand  \f  \varphi
\newcommand \bra {\langle}
\newcommand \ket {\rangle}
\newcommand{\be}{\begin{equation}}
\newcommand{\ee}{\end{equation}}
\newcommand{\ben}{\begin{displaymath}}
\newcommand{\een}{\end{displaymath}}
\newcommand{\ba}{\begin{eqnarray}}
\newcommand{\ea}{\end{eqnarray}}
\newcommand{\ban}{\begin{eqnarray*}}
\newcommand{\ean}{\end{eqnarray*}}
\newcommand{\cro}{\dagger}
\newcommand{\tr}{{\rm Tr}}
\newcommand{\de}{\partial}

\newcommand{\tvet}{\bar{\tau}}

\newcommand{\Frac}[2]{\leavevmode\kern.1em\raise.5ex
\newlength{\www}
\newcommand{\sopra}[2]{\settowidth{\www}{#1}#1\hspace{-\www}#2
\settowidth{\www}{#2}\hspace{-\www}\settowidth{\www}{#1}\hspace{\www}}
\newcommand{\eq}[1]{(\ref{#1})}
\newcommand{\lra}{\leftrightarrow}

\hbox{\the\scriptfont0 #1}
\kern-.1em/\kern-.15em\lower.25ex\hbox{\the\scriptfont0 #2}}
\makeatletter
\newcounter{alphaequation}[equation]
 \def\thealphaequation{\theequation\hbox to
0.6em{\hfil\alph{alphaequation}\hfil}}
\def\eqnsystem#1{
\def\@eqnnum{{\rm (\thealphaequation)}}
\def\@@eqncr{\let\@tempa\relax
\ifcase\@eqcnt \def\@tempa{& & &}
\or \def\@tempa{& &}\or \def\@tempa{&}\fi\@tempa
\if@eqnsw\@eqnnum\refstepcounter{alphaequation}\fi
\global\@eqnswtrue\global\@eqcnt=0\cr}
\refstepcounter{equation}
\let\@currentlabel\theequation
\def\@tempb{#1}
\ifx\@tempb\empty\else\label{#1}\fi
\refstepcounter{alphaequation}
\let\@currentlabel\thealphaequation
\global\@eqnswtrue\global\@eqcnt=0
\tabskip\@centering\let\\=\@eqncr
$$\halign to \displaywidth\bgroup
  \@eqnsel\hskip\@centering
  $\displaystyle\tabskip\z@{##}$&\global\@eqcnt\@ne
  \hskip2\arraycolsep\hfil${##}$\hfil&
  \global\@eqcnt\tw@\hskip2\arraycolsep
  $\displaystyle\tabskip\z@{##}$\hfil
  \tabskip\@centering&\llap{##}\tabskip\z@\cr}

\def\endeqnsystem{\@@eqncr\egroup$$\global\@ignoretrue}
\makeatother

\def\circa#1{\,\raise.3ex\hbox{$#1$\kern-.75em\lower1ex\hbox{$\sim$}}\,}
\newcommand{\mysection}[1]{\setcounter{equation}{0}\section{#1}}
\renewcommand{\theequation}{\thesection.\arabic{equation}}

\def\circa#1{\,\raise.3ex\hbox{$#1$\kern-.75em\lower1ex\hbox{$\sim$}}\,}

\title{
\bf The Effective Lagrangian of the \\
Two Higgs Doublet Model
}

\author{ {\sc P.Ciafaloni}\thanks{ciafalon@ifae.es} \\
         { IFAE, Universitat Aut\`onoma de Barcelona} \\
         { Edifici Cn }\\
         { E-08193 Bellaterra}\\
         \\
         and\\
         \\
         {\sc D. Espriu}\thanks{espriu@greta.ecm.ub.es} \\
         { DECM and IFAE, Universitat de Barcelona} \\
         { Diagonal 647}\\
         { E-08028 Barcelona} \\
         \\ }

\date {}         
\begin{document}
\maketitle

{\Large\begin{abstract}

We consider the two Higgs doublet model extension of the Standard Model
in the limit where all  physical scalar particles are very heavy; too
heavy, in fact, to be experimentally produced in forthcoming experiments.
The symmetry breaking sector can thus be described by an effective chiral
Lagrangian. We obtain the values of the coefficients of the ${\cal O}(p^4)$
operators
relevant to the oblique corrections and investigate to what extent
some non-decoupling effects may remain at low energies.

\end{abstract}}

\vfill
\vbox{
UAB-FT-405\hfil\null\par
UB-ECM-PF-96/23\hfil\null\par
December 1996\hfil\null\par}

\clearpage

\mysection{Introduction}
The two Higgs doublet model (2HDM\cite{2HDM}) is one of the 
most popular extensions
of the Standard Model (SM). It provides a natural way of               
introducing
an additional $U(1)$ (or Peccei-Quinn \cite{PQ}) symmetry, allows for
spontaneous breaking of $CP$ invariance \cite{CPX}, and may provide
an interesting phenomenology of flavour changing neutral currents
\cite{FCNC} ---compatible with the current experimental limits---
if the appropriate form of the scalar potential is chosen. In fact
if Nature has decided that electroweak symmetry breaking should proceed
via elementary scalar fields, it is difficult to answer the question
as to why not two doublets instead of just one as in
the Minimal Standard Model (MSM).

Recent data from LEP \cite{Warsaw96} put stringent limits
on the symmetry breaking sector of the SM. While composite and
QCD-like Technicolor models are not, strictly speaking, ruled out
yet, they are severely constrained. Technicolor groups larger
than $SU(2)_{TC}$ appear extremely unlikely, while the amount
of custodial symmetry breaking in the techni-quark mass sector is
severely limited by the $\rho$ parameter \cite{tclimits}.
Nature seems to be telling us that whatever physics one may think
of adding to the SM, it should, to a good extent, decouple at low energies
(in the technical
sense of Appelquist and Carazzone \cite{AppCar}). Technicolor and similar
theories are non decoupling; finite and calculable corrections to
the low energy parameters $S,T,U$ \cite{Peskin}
or, equivalently, $\varepsilon_1,\varepsilon_2,\varepsilon_3$
\cite{barbieri}
remain even when the mass scale of all new particles is large.
This is why one is able to set severe limits on such theories.
On the contrary, the symmetry breaking sector of the minimal supersymmetric
standard model (MSSM) \cite{susy}
---which contains two Higgs doublets--- is decoupling. In such
theories one can always tune the parameters in such a way
that the additional contribution of this sector to $S,T$ and $U$ is
arbitrarily small (at least in the MSSM, see \cite{LiMa}).
The fact that
the theory is decoupling implies that modifications to the SM results are
small and adjustable.

One may wonder whether the fact that the enlarged symmetry breaking
sector decouples
from low $(\sim M_W)$ energy phenomenology is a generic feature of 2HDM
or it is just limited to the MSSM.
Does the heavy scalar sector decouple from low energy ($\sim M_W$)
phenomenology? And if not, to what extent?
We propose to
investigate this issue here.

Heavy scalar masses usually imply, at least naively,
a strongly interacting symmetry breaking sector thus rendering
the usual linear, perturbative approach questionable.
We shall thus phrase our discussion in the
language of effective Lagrangians.
This technique is the natural one
when all the physical degrees of freedom in
the symmetry breaking sector are heavy, and a separation between light and
heavy degrees of freedom is clear
(which is not the same as saying that the
heavy sector should necessarily decouple, as exemplified by technicolor
models). It should be said right away that
the above situation does not correspond to  the MSSM, where
it is not natural to have all
scalar
fields heavy, and some light scalar must necessarily be
present. Thus our results are not directly applicable to the MSSM.
Rather, our analysis applies to 2HDM
where the masses of all physical scalar particles are very large,
typically somewhere in the TeV region. While 2DHM models with light to
moderate masses
have already been studied \cite{previous}, heavy doublets
do not appear to have been considered in detail, at least to our knowledge.
The symmetry breaking sector just described may or (most likely) may not
correspond to some supersymmetric theory; this does not concern us here.
If additional light fields (such as s-quarks or gauginos) are present we
shall just include them explicitly in our low-energy theory.

We want to keep the light degrees
of freedom only, namely the gauge and Goldstone bosons. The latter are
collected in a unitary matrix $U=\exp{2iG^a T^a/v}$ where $v$ is the
vacuum expectation value that gives the $W$ and $Z$ bosons a mass and
$G_a$ are the Goldstone modes. The matrix $U$ is an element of the
$SU(2)\times SU(2)/SU(2)$
coset space. Given this basic building block and gauge invariance one just
constructs the most general Lagrangian compatible with the desired
symmetries via a derivative expansion; namely \cite{Longhi,ChiralLag}
\be
{\cal L}= {\cal L}^2 + {\cal L}^4+ \ldots
\ee
The indices denote the dimensionality of the corresponding
operators, i.e. two derivatives, four derivatives, etc.
Gauge fields count as one derivative and explicit breaking terms are
forbidden on account of gauge invariance. See \cite{Longhi}
for a classification of all possible operators up to ${\cal O}(p^6)$.
The information on physics beyond the MSM is encoded in the
coefficients of the above effective chiral Lagrangian (ECL).

There are only two independent ${\cal O}(p^2)$ operators
\be
\label{p2}
{\cal L}^2= {v^2\over 4} {\rm Tr}(D_\mu U D^\mu U^\dagger) +
a_0 {v^2\over 4} ({\rm Tr}(\tau_3 U^\dagger D_\mu U))^2
\ee
The first one is universal, its coefficient is fixed by the $W$ mass.
The other one is related to the $\rho$ parameter.
In addition there are a few ${\cal O}(p^4)$ operators with their
corresponding coefficients
\be
\label{p4}
{\cal L}^4={1\over 2} a_1 g g^\prime {\rm Tr}(U B_{\mu \nu} U^\dagger
W^{\mu\nu})
-{1\over 4}a_8g^2{\rm Tr}(U\tau^3 U^\dagger W_{\mu\nu}){\rm Tr}(
U\tau^3 U^\dagger W^{\mu\nu})+ ...
\ee
In the above expression $B_{\mu \nu}$ and $W_{\mu\nu}$ are the
field strength tensors associated to the $SU(2)$ and $U(1)$ gauge fields.
In this paper we shall only consider the self-energy, or oblique,
corrections, which are  dominant in the two Higgs doublet
model just as they are in the MSM. Accordingly,  we shall determine only
those coefficients of the ECL that contribute to two-point functions to 
leading and next-to-leading order in the momentum expansion. 
These consist in just the two operators quoted above after using the 
equations of motion (see however \cite{matias}).

Apart from vacuum polarization effects, the 2HDM
introduces, with respect to the SM, some additional vertex and box
corrections due to the exchange of
scalar particles. Let us phrase the discussion in terms of the familiar
$\varepsilon_1, \varepsilon_2,\varepsilon_3$
parameters \cite{barbieri}.
The experimental bounds on these quantities
are extracted from observables with leptons in the external legs, and
therefore such box and vertex
corrections are, roughly speaking, suppressed by a factor $\sim
(m/M_W \cos\beta)^2$ with respect to typical gauge
corrections\footnote{This holds for the so-called type II models.
For type I models the relevant parameter is $\cot\theta$ instead of
$\tan\beta$; apart from this, the same considerations hold.
See \cite{hunterguide}}.
Here $m$ is the lepton mass and $\tan\beta=v_2/v_1$
depends on the ratio between the two v.e.v's appearing in 2HDM.
With the current experimental limit $\tan\beta<
0.52 M_+$ (GeV) \cite{tanbeta} ($M_+$ is the charged Higgs mass),
and setting for $M_+$ the value $ M_+\approx
600 $ GeV,  we get $\tan\beta \circa{<} 300$. Then the additional box
and vertex contributions due to scalar exchange can be safely neglected
for $e$ and $\mu$ leptons\footnote{Since lepton universality 
is verified by LEP data at the {\it per mille} level, 
we can also include  the $\tau$ lepton in our analysis.}.
The limit of vacuum polarization
dominance is therefore justified and we can compare
our results with the values for
$\varepsilon_i$ extracted from lepton data.
\bigskip

\mysection{The Model and its Non-linear Realization}
{}From the considerations in the previous section it should be clear
that we must, first of all, proceed to the separation of the heavy
and the light degrees of freedom. The subsequent step will be to determine
the actual numerical value of the relevant $a_i$ coefficients,
as a function of the parameters of our underlying theory.
Unfortunately, this last step we can do only within the framework
of perturbation theory which is suspect
in a strongly interacting scalar theory. However, renormalized perturbation
theory turns out to be reliable in a trivial theory,
as the scalar sector
of the MSM appears to be, because the theory is never really strongly
interacting \cite{luscher}.
Yet, this is not necessarily so
in the presence of additional interactions and fields ---particularly
supersymmetry---which may alter the ultraviolet properties of the
theory. We shall thus rely as much as possible on dimensional and power
counting arguments and discuss to what extent these agree with perturbation
theory.

Let us begin by reviewing the model in the usual (weakly coupled)
linear realization. We have
two Higgs doublets
$\phi_1,\phi_2$. For the potential we choose the most general one
respecting $CP$ and the discrete symmetry
$\phi_1\to-\phi_1,\phi_2\to\phi_2$. Imposing this symmetry automatically
avoids an excessive amount of flavour violation \cite{hunterguide}. It
also suppresses spontaneous $CP$ violation.
\be\label{scalarpot}\begin{array}{ccl}
V(\phi_1,\phi_2)&=&\lambda_1(\phi_1^\cro\phi_1-v_1^2)^2+
\lambda_2(\phi_2^\cro\phi_2-v_2^2)^2+
\lambda_3((\phi_2^\cro\phi_2-v_2^2)+(\phi_1^\cro\phi_1-v_1^2))^2+\\
 & &\lambda_4((\phi_1^\cro\phi_1)(\phi_2^\cro\phi_2)-
(\phi_1^\cro\phi_2)(\phi_2^\cro\phi_1))+\lambda_6(
Im (\phi_1^\cro\phi_2))^2 \end{array}\ee
where
\be
\label{vevs}
\phi_1= \left(\begin{array}{c}
\alpha_+\\
\alpha_o \end{array}\right)
\quad
\phi_2= \left(\begin{array}{c}
\beta_+\\
\beta_o \end{array}\right)
\;\;\;\;\;\;\;
\bra\phi_1\ket=\left( \begin{array}{c}
0\\
v_1 \end{array}\right)
\quad
\bra\phi_2\ket=\left( \begin{array}{c}
0\\
v_2  \end{array}\right)
\ee
$v_1,v_2$ are real and $\lambda_i\geq 0$.
We consider the following
$2\times 2$ matrices (as usual $\bar{\phi}=i\tau_2\phi^*$)
\be
\label{phis}
\Phi_{12}=\left(\begin{array}{cc}
\bar{\phi}_1 & \phi_2\end{array}\right)
\;\; \;\;
\Phi_{21}=\left(\begin{array}{cc}
\bar{\phi}_2 & \phi_1\end{array}\right)
\;\; \;\;
\Phi_{21}=\tau_2 \Phi_{12}^*\tau_2
\ee
Under a $SU_L(2)
\times U(1)$ transformation,
$\Phi_{ij}\to \exp[i\bar{\tau}\cdot
\bar{\alpha}]\Phi_{ij}  \exp[-i{\tau}_3\cdot\beta_3]$.
$\Phi_{12}$ and $\Phi_{21}$ transform in fact in the same
way under the larger group $SU_L(2) \times SU_R(2)$, namely $\Phi_{ij}\to
\exp[i\bar{\tau}\cdot\bar{\alpha}]\Phi_{ij}
\exp[-i\bar{\tau}\cdot\bar{\beta}]$. In terms of $\Phi_{ij}$
we define the auxiliary matrices
\be\label{defI}
 I=\Phi_{12}^{\cro}\Phi_{12}\; \; \;
 J=\Phi_{12}^{\cro}\Phi_{21}
\ee
$I$ and $J$ both transform as  $I\to \exp[i\bar{\tau}\cdot
\bar{\beta}] I  \exp[-i\bar{\tau}\cdot\bar{\beta}]$.
Furthermore
\be
\bra I\ket=\frac{v_1^2+v_2^2}{2}+
\frac{v_1^2-v_2^2}{2}\tau_3\; \; \;
\bra J\ket=v_1v_2
\ee
We would like to point out that other parametrizations are possible. For 
instance, we could have embedded the two scalar fields in two 2 x 2 
matrices in this way:  
\be
\Phi_{1}=\left(\begin{array}{cc}
\bar{\phi}_1 & \phi_1\end{array}\right)
\;\; \;\;
\Phi_{2}=\left(\begin{array}{cc}
\bar{\phi}_2 & \phi_2\end{array}\right)
\ee
In this case the $SU(2)\times SU(2)$ symmetry is implemented in a slightly
different way. However, we choose (\ref{phis}) because Goldstone bosons 
fields appear quite naturally in the nonlinear parametrization, as we 
shall see.

In terms of the above matrices the kinetic term reads
\be\label{kinetic}
T(\Phi_{12})=\frac{1}{4}{\rm Tr}(D_\mu\Phi_{12}D^\mu\Phi_{12}^\dagger)
\ee
and the potential can be expressed
as 
\be\label{potential}\begin{array}{ccl}
V(\Phi_{12},\Phi_{21})&=&{\lambda_1\over 4}\{{\rm Tr}[(I-\bra
I\ket)(1+\tau_3)]\}^2+
{\lambda_2\over 4}\{{\rm Tr}[(I-\bra I\ket)(1-\tau_3)]\}^2+ \\
&&\lambda_3\{{\rm Tr}[I-\bra I\ket]\}^2+
{\lambda_4\over 4}{\rm Tr}[I^2-(I\tau_3)^2]+
{\lambda_6\over 4}\{\frac{1}{2i}{\rm Tr}[J-J^\cro]\}^2
\end{array}\ee
This potential is invariant under
$SU_L(2) \times U(1)$, but some terms
are not invariant under
$SU_L(2) \times SU_R(2)$.
They break custodial symmetry and may lead, at least potentially,
to sizeable contributions to the $\rho$ parameter.

After symmetry breaking to $U(1)_{em}$, the matrices $I$ and $J$ get a
v.e.v. and new fluctuations around
the vacuum state appear. Some are massless (the 3 Goldstone bosons)
and other massive (by hypothesis, very massive in our case). We want
to separate these very different degrees of freedom to all
orders in perturbation theory. The massless degrees of freedom will enter
the unitary matrix $U$ and the rest will eventually be integrated out in
the coefficients $a_i$. The problem is somewhat non-trivial. Suppose for
instance that, in analogy to the one doublet case,
we write the 2 x 2 matrices $\Phi_{12}$ and $\Phi_{21}$ as
the product of a unitary matrix and an
hermitian one, e.g.
\be\label{deffi21}
\Phi_{12}={\cal U} H_{12}
\ee
Then the unitary matrix
${\cal U}=\exp(i\theta/v)\exp(i\bar{\tau}\bar{G}/v)$,
where $v^2=(v_1^2+v_2^2)/2$ is the combination of v.e.v.'s
relevant for the $W$ mass,
would hopefully
collect the Goldstone bosons and 
$H_{12}=(\sigma 1+\bar{\tau}\cdot\bar{\gamma})$
would be the extension to the two doublet case of $\sigma 1 $ (notice
the appearance of the additional phase $\theta$ in the 2HDM).
Unfortunately, this separation is way too naive. In fact,
although the $\bar{G}$ fields
do not appear in the scalar potential, they mix with the fields in $H$
due to the kinetic term
and therefore cannot be identified as Goldstone bosons.

Yet the above decomposition is quite suggestive because when
one substitutes back in the scalar potential any decomposition
of the form $\Phi_{12} = U M_{12}$, where $M_{12}$ is not
necessarily hermitian,
$U$ drops from the potential exactly. Thus instead
of assuming that $M$ is hermitian, we shall allow for a more general
matrix (this is just fine, as long as the decomposition is
still unique). We single out the Goldstone bosons by making
an infinitesimal gauge transformation specialized to the broken generators:
\be\label{gauge}
\delta_\epsilon =
1+i\bar{T^L}\cdot\bar{\epsilon}^L+iT^R\epsilon^R=
1+iT_+^LG_+ +iT_-^LG_-+i\frac{T_3^L-T_3^R}{2}G_0
\ee
Acting with such a transformation on the vacuum configuration for e.g.
$\Phi_{12}$ we obtain
\be\label{fi12}
\delta_\epsilon\;\Phi_{12}=\left(\begin{array}{cc}
v_1+iG_0\frac{v_1}{v}&
i\sqrt{2}G_+\frac{v_2}{v}      \\
i\sqrt{2}G_-\frac{v_1}{v} &
v_2-iG_0\frac{v_2}{v}\end{array}\right)
\ee
And an analogous expression for $\Phi_{21}$.
Goldstone bosons and massive excitations must be orthonormal for the
kinetic terms to be  diagonal. Once the
former have been identified, the latter are uniquely determined. We obtain
\be\label{fi12li}
\Phi_{12}=\left(\begin{array}{cc}
Re[\alpha_o]+i(G_0\frac{v_1}{v}+A_0\frac{v_2}{v})&
\sqrt{2}(H_+\frac{v_1}{v}+iG_+\frac{v_2}{v} )       \\
\sqrt{2}(H_-\frac{v_2}{v}+iG_-\frac{v_1}{v}) &
Re[\beta_0]+i(-G_0\frac{v_2}{v}+A_0\frac{v_1}{v})\end{array}\right)
\ee
Keeping terms at most linear in the fields $\Phi_{12}$ can also be
written  as
\be\label{fi12nonli}
\Phi_{12}=\exp[i\frac{\bar{G}\cdot\tvet}{v}]\left(
    \begin{array}{cc}
    Re[\alpha_0]+iA_0\frac{v_2}{v}&
    \sqrt{2}H_+\frac{v_1}{v}       \\
    \sqrt{2}H_-\frac{v_2}{v} &
    Re[\beta_0]+iA_0\frac{v_1}{v}
    \end{array}\right)
\ee
Notice that an alternative form for $\Phi_{12}$, useful
for calculations, is
\be \Phi_{12}=\exp[i\frac{\bar{G}\cdot\tvet}{v}] (\sigma+iA_0+\tvet\bar{H})
  \left(
  \begin{array}{cc}
    \frac{v_2}{v}&
     0      \\
     0           &
    \frac{v_1}{v}
    \end{array}
   \right)
\ee
with
\be
\sigma=\frac{1}{2\sqrt{2}}(\frac{Re[\alpha_0]}{\sin\beta}
+\frac{Re[\beta_0]}{\cos\beta})\;\;\;
\;\;\;
H_3=\frac{1}{2\sqrt{2}}(\frac{Re[\alpha_0]}{\sin\beta}
-\frac{Re[\beta_0]}{\cos\beta})
\ee
This is the expression we were after. It satisfies the following properties:
1) It is a parametrization of $\Phi_{12}$; 2) It is of the form
$\Phi_{12}=U M_{12}$, where $U\in SU(2)$;
3) It diagonalizes the kinetic terms by construction; 4) It
can be proven to be unique.

With this parametrization we also have
\be
\Phi_{21}=\tau_2\Phi^*_{12}\tau_2=UM_{21}=
\exp[i\frac{\bar{G}\cdot\tvet}{v}]
(\sigma-iA_0-\tvet\bar{H})
\left(\begin{array}{cc}
    \frac{v_1}{v}&
     0      \\
     0           &
    \frac{v_2}{v}
    \end{array}\right)
\ee
We now plug the above decomposition $\Phi_{12}=UM_{12}$
into the kinetic term (\ref{kinetic})
\be
T=
\frac{1}{4} {\rm Tr}[D_\mu(UM_{12})
(D_\mu(UM_{12}))^\dagger]
\ee
$M_{21}$ does not appear here, but it does in the potential terms.

Naively, setting the masses of all heavy particles to infinity would
take us to the minimum of the potential, $M_{12}=\bra M_{12}\ket$.
Plugging this back in (\ref{kinetic}), and using 
$\tr(\tau_3 D_\mu U D^\mu U^\dagger)=0$ we recover at the classical
level the
lowest dimensional term  $v^2 {\rm Tr}(D_\mu U D^\mu U^\dagger)$
in the ECL.
Letting
$M_{12}\to\bra M_{12}\ket$ and $D_\mu\to\de_\mu$ we obtain
the kinetic terms
\be\label{kin}
{1\over v^2}\de_\mu \bar{G} \de^\mu\bar{G}\tr(\bra M_{12}\ket
\bra M_{12}
\ket^\cro)+
\tr(\de_\mu M_{12}\de^\mu M_{12}^\cro)
\ee
The crossed term vanishes
\be
{i\over v}\tr(\tvet\de_\mu\bar{G}
(\bra M_{12}\ket\de^\mu M_{12}^\cro-\de^\mu M_{12}\bra M_{12}\ket^\cro))=
2{{v_1 v_2}\over v^2}\tr(\de_\mu\bar{G}\tvet\de^\mu A_0)=0
\ee
We are then left with the canonical diagonal kinetic terms for the fields
$\bar{G},A_0,H_1^0,H_2^0,H_+,H_-$.  The  fields
$H_1^0$ and $H_2^0$ are the
mass eigenvalues of the mass matrix in the
$CP$ even neutral sector.
This leads to the appearance of the mixing angle $\alpha$
which depends on the parameters of the potential
\be
\label{h1h2}
H_1^0=\frac{Re[\alpha_o]}{\sqrt{2}}\cos\alpha+
\frac{Re[\beta_o]}{\sqrt{2}}\sin\alpha \qquad
H_2^0=-\frac{Re[\alpha_o]}{\sqrt{2}}\sin\alpha+
\frac{Re[\beta_o]}{\sqrt{2}}\cos\alpha
\ee
Neither $H_1^0$ nor $H_2^0$ have simple
interaction terms. In particular both have non vanishing v.e.v.'s.  The
combinations
\be
S= \sin(\alpha-\beta) H_1^0 + \cos(\alpha-\beta) H_2^0 \qquad
H= \cos(\alpha-\beta) H_1^0 - \sin(\alpha-\beta) H_2^0
\ee
on the other hand
``diagonalize" the interaction pieces in the Lagrangian.
By this we mean that the field $H$ (which has $\bra H\ket=v$) has
exactly the same interaction terms as the standard Higgs would have,
in particular the coupling $g M_W H W^+ W^-$, characteristic of a
spontaneously broken theory
(yet $H$ is not an eigenstate of the mass matrix, as mentioned). On the
other hand the field $S$ (with $\bra S\ket=0$) 
does not have any couplings of the above
form. This observation turns out to be important to understand our results.

Although classically  $a_0$, $a_1$ and $a_8$ vanish in the
limit where all scalar particles are very massive, at the quantum level these
coefficients
will be generically non-zero. To obtain their true value we must integrate
out the fields contained in the matrices $M_{12}$ and $M_{21}$.

Since it accompanies a custodially breaking operator,
$a_0$ must on symmetry grounds
be proportional to a typical
mass splitting or a custodial breaking parameter, such as ${g^\prime}^2$.
Naively,
\be
\label{naive}
a_0\sim \{ {g^2\over {16\pi^2}}{\Delta M^2 \over M_W^2},
{{g^\prime}^2\over {16\pi^2}} \log {M_s^2\over M_W^2}\}
\ee
where $M_s$ is a typical heavy-scalar mass.
In fact we will see that the dependence on the quadratic mass splittings,
$\Delta M^2=M^2-M_s^2$,
is quadratic and not linear. At any rate
non-decoupling effects may be important if large mass splittings are
present. (As is well known, the Appelquist-Carazzone decoupling theorem
does not go through for spontaneously broken theories \cite{subtlety}.)
Other potential non-decoupling effects are contained in the coefficients
of the ${\cal O}(p^4)$ operators. On dimensional grounds these
coefficients will be of order
\be
a_i\sim {1\over {16\pi^2}}( \log\frac{M_s^2}{M_W^2} + c + {\cal O}(
\frac{\Delta M^2}{M_s^2})) \qquad (i\neq 0)
\ee
$c$ is a finite constant and
$M_s$ is the mass of the  $CP$-odd scalar,
taken as reference scale.
These terms will be less important at low energies since  it is clear
that the leading pieces in the  momentum
expansion are contained in the $d=2$ terms. At energies
$q^2 \ll 16\pi^2 v^2$
the ${\cal O}(p^4)$ operators will
be suppressed with respect to the ${\cal O}(p^2)$ ones, although
they
rapidly become important as the energy increases.

At this point we should re-examine the field contents of our theory in the
non-linear realization. We have three Goldstone bosons collected in the
unitary matrix $U$, two charged Higgses $H_+$ and $H_-$, two $CP$-even
neutral fields $\sigma$ and $H_3$, or equivalently $H_1^0$, $H_2^0$, and
the $CP$-odd neutral $A_0$. All of them are supposed to be heavy.

\mysection{The Effective Lagrangian Coefficients}
The part of the ${\cal O}(p^4)$ effective Lagrangian
relevant for the determination of the oblique corrections is
the one given in eq. (\ref{p4}).
The corresponding coefficients
contain the traces of the underlying theory
accessible at experiments much below the energy scales $M_s$
or $4\pi v$, whichever is smaller.

It is customary to parametrize possible departures from the
MSM with the parameters $\varepsilon_1$,$\varepsilon_2$ and
$\varepsilon_3$.\cite{barbieri}
These parameters, in the limit of vacuum
polarization dominance, coincide with the $e_i$-parameters defined by
\begin{eqnarray}\label{eq:e1e2e3}
\frac{1}{M_W^2}\big(A_{33}(0)-A_{WW}(0)\big)&\equiv& e_1\nonumber \\
F_{WW}(M_W^2) - F_{33}(M_W^2) &\equiv& e_2\\
\frac{c}{s} F_{30}(M_Z^2)&\equiv& e_3 \nonumber
\end{eqnarray}
in terms of the vacuum polarization amplitudes
\be
\Pi_{\mu\nu}^{ij}(q)=-i g_{\mu\nu}[A^{ij}(0)+ q^2 F^{ij}(q^2)]+
q_\mu q_\nu~{\rm terms}
\ee
where $i,j=W,0,3$ stand for the $W^{\pm},B$ and $W^3$ gauge boson fields,
respectively.
The $S,T$ and $U$ parameters \cite{Peskin} are trivially related to
the above.

In an effective theory such as the one described by the Lagrangian
(\ref{p2}) and (\ref{p4}) $\varepsilon_1$,$\varepsilon_2$
and $\varepsilon_3$ receive one loop
contributions from the leading ${\cal O}(p^2)$ term $v^2{\rm Tr}(D_\mu
UD^\mu U^\dagger)$ and tree level contributions from
the $a_i$. Thus
\be
\varepsilon_1= 2 a_0+\ldots \qquad  \varepsilon_2= -g^2a_8 +\ldots
\qquad \varepsilon_3= -g^2a_1+\ldots
\ee
where the dots symbolize the one-loop
${\cal O}(p^2)$ contributions. The latter
are totally independent of the specific symmetry breaking sector.
The non-linear nature of ${\cal L}^2$ induces new divergences
which are absorbed by a proper redefinition of the $a_i$. These divergences
are by construction independent of the underlying theory, so we know the
logarithmic dependence of the coefficients for free for any two
Higgs doublet model \cite{Longhi,ChiralLag}
\be
a_0\sim {{g^\prime}^2\over 16\pi^2}{3\over 8}\log {M_W^2\over M_s^2}
\ee
\be
a_1\sim {1\over {16\pi^2}} {1\over 12} \log {M_W^2\over M_s^2}
\ee
\be
a_8\sim 0
\ee
By construction this logarithmic dependence is exact, even in
the non-perturbative large $M_s$ limit. The above are
renormalized coefficients in the $\overline{MS}$ scheme.

The coefficients $a_i$ contain in addition constant
and ${\cal O}(\Delta
M^2/M_s^2)$ corrections. These subleading contributions
are non-universal and have to be determined by matching
e.g. the renormalized self energies,
or an appropriate combination thereof,
between fundamental and effective theories.
For instance, we can match the combinations of self energies appearing
in (\ref{eq:e1e2e3}). When we compute the values of the $a_i$
coefficients via the matching conditions, most of the diagrams
cancel between both sides of the matching equation. Only those
containing at least one heavy particle contribute ---properly expanded in
$p^2$---
 to the
coefficients of the ECL.

The values of the $\varepsilon_i$ in a 2HDM  have already been calculated
in the past \cite{previous} in the linear perturbative regime.
In the non-linear
realization there are differences already at the level
of Feynman rules,
and some simplifications worth  pointing out.
For instance the vertex with one scalar neutral Higgs
boson $H_3$, one charged Goldstone boson $G_+$ and one gauge boson
$W_\mu^-$, is given by
$g H_3\de_\mu G_+W_\mu^-$ in the nonlinear case, and by
$\frac{g}{2} H_3\stackrel{\leftrightarrow}{\de_\mu}G_+ W_\mu^-$
in the linear parametrization.

\begin{figure}
\setlength{\unitlength}{1cm}
\begin{center}
\begin{picture}(17,19)
\put(2.5,16.4){(a)}
\put(7,11){(b)}
\put(7,5){(c)}
\put(-3.6,-7){\includegraphics{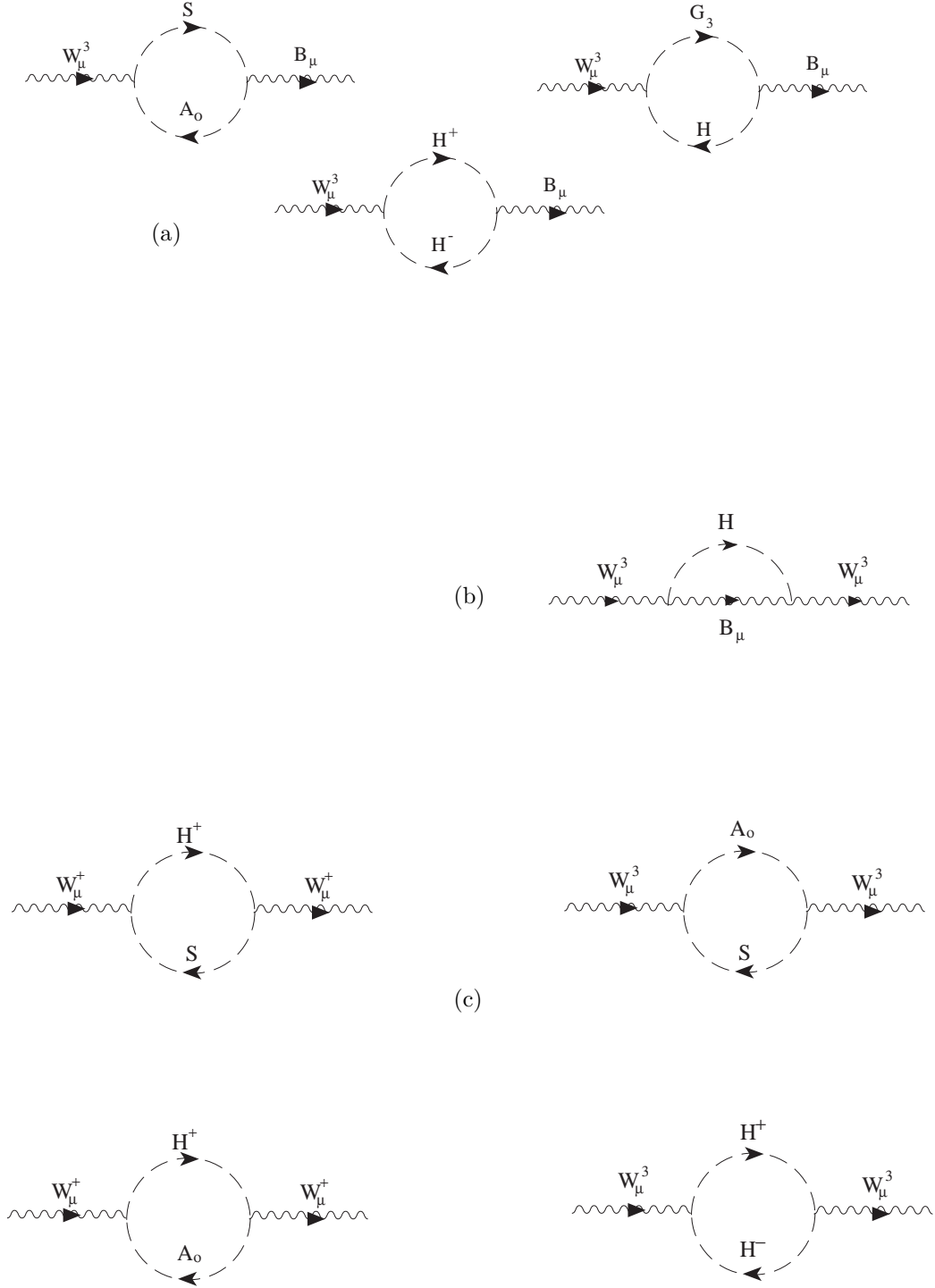}}
\end{picture}
\end{center}
\caption{
Feynman diagrams required to evaluate the coefficients $a_i$
via matching conditions for
 $\varepsilon_i $. Diagrams (a) are for  $\varepsilon_3$, diagrams (c) for 
$\varepsilon_2$ and (diagrams (c) + diagram (b)) for $\varepsilon_1 $.
\label{fig:e1e2}
        }
\end{figure}

We are interested only in the leading corrections in the limit
$q^2\approx M_W^2\ll M_s$, where $M_s$ is a typical heavy
scalar mass.
Then we can set $M_W=M_G=0$ ($M_G$ is the gauge-dependent
Goldstone bosons mass) in the internal lines. Moreover,
when calculating the diagrams contributing to $e_2$ and $e_3$, those with
gauge bosons in the internal lines do not contribute.
These simplifications follow from simple dimensional considerations. For
instance, the diagrams containing one internal vector boson line are
proportional to $g^2 M_W^2$  and their overall contribution to $e_3$
is ultraviolet finite.
Since $e_3$ is proportional to the dimensionless derivatives
of the vacuum polarization,
there must be a $M_s^2$ in the denominator, and so the contribution is
proportional to $M_W^2/M_s^2$ and therefore subleading.
In the same way it is possible to see
by inspection that some diagrams (such as tadpoles) do not
contribute in the matching relations for  $e_1$.
We are left with the diagrams in fig. \ref{fig:e1e2} for
the matching relations for $e_1$,
$e_2$ and $e_3$.
The only nonzero masses in these diagrams are the heavy scalar masses.

If we neglect terms of ${\cal O}(M_W^2/M_s^2)$ the values
one obtains from (A.3) for the $a_i$ coefficients are
\be
\label{a0}
a_0=   {g^2\over 16\pi^2}{1\over 24}{{\Delta_+^2
         (\Delta_+^2-\sss \Delta_1^2-\ccc \Delta_2^2)}\over{M^2_sM_W^2}}
           +{{g^\prime}^2\over 16\pi^2}{3\over 8}(\log {M_W^2\over M_s^2}
           +{5\over 6}+ {\cal O}({\Delta_i^2\over M_s^2}))
\ee
\be
\label{a1}
a_1=  {1\over {16\pi^2}} {1\over 12} (\log {M_W^2\over M_s^2}
           +{5\over 6}+ {\cal O}({\Delta_i^2\over M_s^2}))
\ee
\be
\label{a8}
a_8=  {\cal O}({1\over {16\pi^2}} {{\Delta_i^2}\over M_s^2})
\ee
These expressions are valid in the limit $M_W^2\le\Delta_i^2\ll M_s^2$.
Here $\sss$
and $\ccc$ are the sinus and cosinus of $\alpha-\beta$, where $\alpha$
is the angle that mixes the two neutral scalars
and $\tan\beta=v_2/v_1$. The quantities $\Delta_i^2$ denote the quadratic
mass splittings of the scalar labelled $i$ ($i=+,-$ correspond to the
charged Higgses, $i=1,2$ to the two $CP$-even neutrals) with respect
to the $CP$ odd neutral scalar, whose mass is taken as reference scale;
that is $\Delta_i^2\equiv M_i^2-M_s^2$. In the above expressions we have
kept the dominant terms only.

The first observation is that the constant pieces in
(\ref{a0}), (\ref{a1}) and (\ref{a8}) are
the same ones as those obtained in {\cite{matias,herrero,nyffler}}
in a similar
calculation in the MSM, with the obvious replacement $M_s\to M_H$.
Some non-decoupling effects do remain in the 2HDM, but they are 
identical to the MSM ones.
Thus in the limit where the mass splittings are negligible compared
to the typical scale in the symmetry breaking sector the two Higgs
doublet model cannot be distinguished by any low energy experiment
from the minimal Standard Model.
For instance the combination of coefficients
\be
-{2\over 9} \cos^2\theta_W a_0 + g^2 \sin^2\theta_W a_1
+g^2 \cos^2\theta_W a_8
\ee
which is zero in the MSM in the limit of a heavy Higgs mass, is
still zero in the 2HDM provided that $\Delta_i^2 \ll M_s^2$. (The
above combination corresponds to straight lines in the
$\Gamma_l$, $\sin_{eff}^2\theta_W$ plane. See \cite{matias}.)

Can we understand this? As discussed in \cite{subtlety,nyffler}, the source
of the non-decoupling effects can be traced back to the appearance
of mixed heavy/light vertices without derivative couplings.
These are characteristic
of spontaneously broken theories (in which one shifts some field).
An archetypical example has already been mentioned: the vertex
$g M_W H W^+ W^-$, but also vertices such as $g H G^+ W^-$.
The point is that, once we write the neutral scalar sector in terms of the
$H$ and $S$ fields, these couplings are identical to the ones in
the MSM.  $H$ is the only culprit of non-decoupling effects, while
fields such as $S$, $H_+$, $H_-$ and $A_0$ leave, after integrating them
out, contributions suppressed by powers of heavy masses, i.e.
they decouple.

A second remark is that the dependence of $a_0$ on the
quadratic mass splittings is of order
$g^2 (\Delta_i^2)^2/M_W^2M_s^2$ or $g^{\prime 2} (\Delta_i^2)/M_s^2$ .
Once we understand that non decoupling effects must be the same in the
limit $\Delta_i^2/M_s^2\to 0$ as in the MSM, it is clear that
the naive counting in (\ref{naive}) cannot hold. We need
an additional power of $M_s^2$ in the denominator, and that
forces $\Delta_i^2$ to appear quadratically.

What happens beyond perturbation theory? This we cannot answer
precisely of course. But we can stand by the order of magnitude
estimates derived from the general arguments given above. We cannot
prove that the constant pieces in $a_1$ for instance will
be the same after a non-perturbative calculation. However, we
can certainly conjecture that the non-decoupling pieces (those
not suppressed by inverse powers of $M_s^2$) will be the same as
in the MSM. And for the latter, perturbation theory turns out to be
eventually reliable as previously discussed.

Some non-standard non-decoupling effects remain however in the case where
the mass splittings are sizeable, $\Delta_i^2\sim M_s^2$. To discuss to
what extent these effects are visible it is best to return to the
$\varepsilon_i$ parameters.

\mysection{Comparison with Experiment and Conclusions}
We now compare the results on $\varepsilon_i$-parameters with the
experimental data from LEP and SLD. Our aim
is to restrict the allowed parameter space in the sort of models we are
considering.  If we set a maximum
value $\Delta^2$ for all the quadratic mass splittings,
$|\Delta_i^2|<{\Delta^2}$,
we obtain from (\ref{eq:eiappr}) that
the $\varepsilon_i$ parameters lie between two
extreme values given below.
To ease the comparison with the MSM we quote below the deviations
with respect to the values for $\varepsilon_i$ obtained there.
This is the reason why the value of $M_H$ (the Higgs mass in the
MSM) appears in the expressions.
\begin{eqnsystem}{eq:eimaxappr}
\delta\varepsilon_1^{max}&\approx&-\frac{3}{4}\frac{g^{\prime 2}}{16\pi^2}
\log\frac{M_s^2}{M_H^2}+ \frac{1}{16\pi^2}
\frac{{\Delta^2}}{M_s^2}(\frac{g^2}{6}\frac{{\Delta^2}}{M_W^2}+\frac{3}{4}
g^{\prime 2} +\frac{3}{8}g^{\prime 2}\frac{{\Delta^2}}{M_s^2}) \\
\delta\varepsilon_1^{min}&\approx&-\frac{3}{4}\frac{g^{\prime 2}}{16\pi^2}
\log\frac{M_s^2}{M_H^2}-
\frac{1}{16\pi^2}\frac{{\Delta^2}}{M_s^2}(\frac{g^2}{48}
\frac{{\Delta^2}}{M_W^2}+\frac{3}{4}
g^{\prime 2} -\frac{3}{8}g^{\prime 2}\frac{{\Delta^2}}{M_s^2})\\
\delta\varepsilon_2^{max}&\approx&
\frac{g^2}{16\pi^2}\frac{1}{240}(\frac{{\Delta^2}}{M_s^2})^2\\
\delta\varepsilon_2^{min}&\approx&
-\frac{g^2}{16\pi^2}\frac{1}{30}(\frac{{\Delta^2}}{M_s^2})^2\\
\delta\varepsilon_3^{max}&\approx&
\frac{g^2}{16\pi^2}\frac{1}{12}\log\frac{M_s^2}{M_H^2}
+\frac{g^2}{16\pi^2}\frac{5}{24}\frac{{\Delta^2}}{M_s^2}
-\frac{g^2}{16\pi^2}\frac{1}{80}(\frac{{\Delta^2}}{M_s^2})^2\\
\delta\varepsilon_3^{min}&\approx&
-\frac{g^2}{16\pi^2}\frac{1}{12}\frac{{\Delta^2}}{M_s^2}+
\frac{g^2}{16\pi^2}\frac{1}{24}(\frac{{\Delta^2}}{M_s^2})^2
\end{eqnsystem}
Expressions (\ref{eq:eimaxappr}a-d) are the
exact maxima and minima of the corresponding expressions given in
(\ref{eq:eiappr}). The
expressions for $\delta\varepsilon_3^{max,min}$ are, however,
simplified approximations to the corresponding  maximum and minimum of
(\ref{eq:eiappr}).
Terms of order $(\Delta^2/M_s^2)^3$ and
higher have been neglected in all cases. 
Note that these expressions do not depend at
all on $\sin (\alpha-\beta)$. Moreover, we checked
that they differ from the actual minimum obtained from eqs.
(\ref{deltaeps}) by less than 5 $\%$ if the mass splittings are less than
150 GeV.

\begin{figure}
\setlength{\unitlength}{1cm}
\begin{center}
\begin{picture}(17,8.3)
\put(0,0){\includegraphics{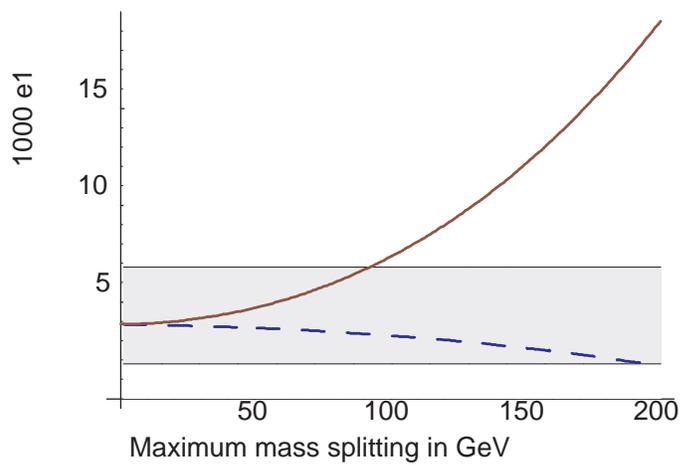}}
\end{picture}
\end{center}
\caption{Maximum (continuous line) and minimum (dashed line)
possible values for
$10^3 \times \varepsilon_1$ in the 2HDM as a
 function of the largest scalar linear mass splitting $\Delta_{max}$.
The grey zone corresponds to the allowed
experimental values\label{fig:eps1} }
\end{figure}

The extraction of the $\varepsilon_i$ parameters from experiment can be
done e.g. along the lines of
\cite{caravals}. Using the latest experimental data
from LEP \cite{Warsaw96}, we obtain \cite{alta}
\begin{eqnarray}
\varepsilon_1=(4.7 \pm 1.3   ) \;\times\; 10^{-3}\nonumber\\
\varepsilon_2=(-7.8  \pm 3.3  ) \;\times\; 10^{-3}\\
\varepsilon_3=(4.8  \pm 1.4  ) \;\times\; 10^{-3}\nonumber
\end{eqnarray}
\frenchspacing
$\varepsilon_1$ is the most restrictive parameter due to its strong
dependence on the splittings and we focus our attention on it.
In fig. \ref{fig:eps1} we plot maximum and minimum values values  for
$\varepsilon_1$
in function of the maximum allowed {\it linear} mass splitting
$\Delta_{max}\approx \Delta^2/2M_s$ (that is,
$|M_i-M_s|<\Delta_{max} $ for all i). 
These values for $\varepsilon_1$ are obtained by adding the MSM 
contribution as taken from $\cite{caravals}$ and
our expressions (\ref{eq:eimaxappr}a,b). A value for $M_H$ in the MSM of
300 GeV was chosen, but the graph itself is
of course independent of the particular value of $M_H$ one chooses.
Also,  the experimentally allowed
values for $\varepsilon_1$ are shown;
it is easy to see that splittings of the order of 100 GeV
or more around a reference
mass of approximately  600 GeV are perfectly allowed.

In conclusion, we have analyzed the situation in which the
symmetry breaking sector of the SM consists in two scalar
doublets with masses in the TeV region. We have separated
the light and heavy degrees of freedom and constructed
an effective chiral Lagrangian for the former. The information
about the latter is contained in a few low energy coefficients.
We have shown that these coefficients can be calculated in terms of a few
Feynman diagrams (see fig. \ref{fig:e1e2}).
We have found that the models exhibits non decoupling effects;
that is, non zero values for the coefficients of operators
with dimensionality $d\le 4$ even in the $M_s\to \infty$ limit.
These non decoupling effects in the limit of exact custodial
symmetry are exactly the same ones as in the MSM.
We have analyzed which restrictions current data set
on two Higgs doublet models; due to the equivalence
between the two Higgs doublet model and the MSM in the
limit of exact custodial symmetry in the scalar potential
the current bounds are very weak.

\bigskip\bigskip
\noindent
{\bf ACKNOWLEDGMENTS \hfil}

\noindent
We would like to thank M.J.Herrero for discussions
concerning possible non-decoupling effects in supersymmetric
theories, and V.Koulovassilopoulos for several comments regarding the
manuscript.
P.C. acknowledges a fellowship from INFN-Frascati-Italy.
This research has been partially supported by grants
ERB-CHRX-CT930343 (E.U.), AEN95-0590 (CICYT) and
GRQ93-1047 (CIRIT).

\nonfrenchspacing

\appendix\mysection{Formulae for $\varepsilon_i$ parameters}
We present here the exact result in the 2HDM to the $\delta\varepsilon_i$
parameters at one loop, defined as $\delta \varepsilon_i=
\varepsilon_i^{2HDM}-\varepsilon_i^{MSM}$
\begin{eqnarray}\label{deltaeps}
\delta\varepsilon_1&=&\frac{g^2}{16\pi^2}\frac{1}{4 M_w^2}
[\sss f(M_{H^+}^2,M_{H_2^o}^2)+
\ccc f(M_{H^+}^2,M_{H_1^o}^2)+f(M_{H^+}^2,M_{A^o}^2)\nonumber \\
&&
-\sss f(M_{A^o}^2,M_{H_2^o}^2)-\ccc f(M_{A^o}^2,M_{H_1^o}^2)] 
-\frac{g^{\prime 2}}{16\pi^2}\frac{3}{4}(\sss\log[
\frac{M_{H_2^o}^2}{M_H^2}]
+ \ccc\log[\frac{M_{H_1^o}^2}{M_H^2}])\nonumber \\
\delta\varepsilon_2&=&-\frac{g^2}{16\pi^2}\frac{1}{12}
[\sss g(M_{H^+},M_{H_2^o})+
\ccc g(M_{H^+}^2,M_{H_1^o}^2)+g(M_{H^+}^2,M_{A^o}^2)\\
&&-\sss g(M_{A^o}^2,M_{H_2^o}^2)-\ccc g(M_{A^o}^2,M_{H_1^o}^2)]\nonumber \\
\delta\varepsilon_3&=&\frac{g^2}{16\pi^2}\frac{1}{12}
[\sss  g(M_{A^o},M_{H_1^o})+\ccc  g(M_{A^o},M_{H_2^o})\nonumber \\
&&+\frac{1}{2}\log[\frac{M_{A^o}^2}{M_H^2}]
-\log[\frac{M_{H^+}^2}{M_H^2}]+(\frac{1+\ccc}{2})
\log[\frac{M_{H^o_1}^2}{M_H^2}]+(\frac{1+\sss}{2})
\log[\frac{M_{H^o_2}^2}{M_H^2}]]\nonumber
\end{eqnarray}
where
\begin{eqnsystem}{eq:fg}
f(a,b)&\equiv& \frac{ab}{a-b}\log[\frac{b}{a}]+\frac{a+b}{2}\\
g(a,b)&\equiv& -\frac{5}{6}+\frac{2ab}{(a-b)^2}+
\frac{(a+b)(a^2-4ab+b^2)}{2(a-b)^3}\log\frac{a}{b}
\end{eqnsystem}
Here $\sss\equiv\sin^2(\alpha-\beta)$  where $\alpha$
is the angle that mixes the two neutral scalars, and
$\tan\beta=\frac{v_2}{v_1}$. $M_H$ is the MSM Higgs mass.
These results coincide with the ones previously published \cite{previous}.
We obtained them in the Feynman-t'Hooft gauge with linear gauge fixing
\cite{matias}.
Since we are interested in the large scalar masses limit,
we set a reference mass $M_s=M_{A^0}$, assumed large,  and
expand functions (\ref{eq:fg}) in the quadratic mass splittings. Defining
$\Delta_i^2\equiv M_i^2-M_s^2$, we obtain
\begin{eqnarray}\label{eq:eiappr}
\delta\varepsilon_1&\approx&\frac{g^2}{16\pi^2}\frac{1}{12}
\frac{\Delta_+^2(\Delta_+^2-\sss \Delta_1^2-\ccc \Delta_2^2)}{M^2_sM_w^2}
\nonumber\\
&&-\frac{3}{4}\frac{g^{\prime 2}}{16\pi^2}\log\frac{M_s^2}{M_H^2}
-\frac{3}{4}\frac{g^{\prime 2}}{16\pi^2}\ccc\log(1+
\frac{\Delta_1^2}{M_s^2})
-\frac{3}{4}\frac{g^{\prime 2}}{16\pi^2}\sss\log(1+
\frac{\Delta_2^2}{M_s^2})  \nonumber\\
\delta\varepsilon_2&\approx&-\frac{g^2}{16\pi^2}\frac{1}{60}
\frac{\Delta_+^2(\Delta_+^2-\sss \Delta_1^2-\ccc \Delta_2^2)}{M_s^4}
\\
\delta\varepsilon_3&\approx&\frac{g^2}{16\pi^2}\frac{1}{120}
\frac{\sss (\Delta_1^2)^2+\ccc (\Delta_2^2)^2}{M_s^4}\nonumber\\
&&+\frac{g^2}{16\pi^2}\frac{1}{12}[\log\frac{M_s^2}{M_H^2}+
\frac{1+\ccc}{2}\log(1+\frac{\Delta_1^2}{M_s^2})+
\frac{1+\sss}{2}\log(1+\frac{\Delta_2^2}{M_s^2})-
\log(1+\frac{\Delta_+^2}{M_s^2})] \nonumber
\end{eqnarray}

\end{document}